\documentclass[]{spie}  

\usepackage{amsmath,amsfonts,amssymb}
\usepackage{graphicx}
\usepackage[colorlinks=true, allcolors=blue]{hyperref}
\usepackage{subcaption}
\usepackage{makecell}
\usepackage{adjustbox}

\title{Weakly-Supervised Detection of Bone Lesions in CT}

\author[a]{Tao Sheng}
\author[a]{Tejas Sudharshan Mathai}
\author[b]{Alexander Shieh}
\author[a]{Ronald M. Summers}
\affil[a]{Imaging Biomarkers and Computer-Aided Diagnosis Laboratory, Radiology and Imaging Sciences, Clinical Center, National Institutes of Health, Bethesda, USA}
\affil[b]{Departments of Interventional Radiology and Imaging Physics, University of Texas MD Anderson Cancer Center, Houston, USA}

\authorinfo{Send correspondence to T.S.M: tejas dot mathai at nih dot gov}

\begin{document} 
\maketitle

\begin{abstract}

The skeletal region is one of the common sites of metastatic spread of cancer in the breast and prostate. CT is routinely used to measure the size of lesions in the bones. However, they can be difficult to spot due to the wide variations in their sizes, shapes, and appearances. Precise localization of such lesions would enable reliable tracking of interval changes (growth, shrinkage, or unchanged status). To that end, an automated technique to detect bone lesions is highly desirable. In this pilot work, we developed a pipeline to detect bone lesions (lytic, blastic, and mixed) in CT volumes via a proxy segmentation task. First, we used the bone lesions that were prospectively marked by radiologists in a few 2D slices of CT volumes and converted them into weak 3D segmentation masks. Then, we trained a 3D full-resolution nnUNet model using these weak 3D annotations to segment the lesions and thereby detected them. Our automated method detected bone lesions in CT with a precision of 96.7\% and recall of 47.3\% despite the use of incomplete and partial training data. To the best of our knowledge, we are the first to attempt the direct detection of bone lesions in CT via a proxy segmentation task.

\end{abstract}


\section{INTRODUCTION}
\label{sec:intro}  

Bones are common destinations for the metastatic spread of cancer from the breast or prostate \cite{Macedo2017}. Bone metastases are usually associated with poor prognosis, reduced quality of life, and various co-morbidities, including an increased risk of fractures, severe pain, or amputation \cite{Macedo2017}. Bone lesions can be categorized as: osteolytic (bone destruction), osteoblastic (bone formation), or mixed lytic and blastic, regardless of whether they are due to a primary tumor, metastases, or incidental findings. Benign hemangiomas and Schmorl's nodes are also often seen, while common observations that are not typically considered ``lesions'' are also observed, such as degenerative changes (osteophytes), degenerative disc diseases, osteoporosis, and osteoarthritis. 

CT imaging is the preferred modality of choice to determine the extent of metastatic spread. Radiologists routinely scroll through a CT volume, identify, and measure the sizes of ``significant'' findings \cite{Eisenhauer2009} in a few slices of the CT volume. However, bone lesions can be difficult to spot due to the myriad variations in their sizes (many are $<$ 1cm), shapes (symmetric or asymmetric), and appearances (hyper- or hypo-intense, or both at once). Moreover, radiologists do not annotate all findings in all slices as the manual measurement is time-consuming and cumbersome. There have been many approaches in literature \cite{Faghani2023, Noguchi2022, Yao2017_mixedSpineMetsDetPETCT, Yao2015_boneMetsCT, Yinong2016_osteophytesPETCTCNN, Burns2016_vertebralBodyFracturesCT, Roth2015_scleroticSpineMetsCTCNN, Burns2013_scleroticMetsCTSVM, Liu2014_epiduralMassesCT, Yao2012_scleroticRibMetsCT, Wiese2012_scleroticBoneMetsCT, Shieh2023_ULD, Frazier2023_ULD, Naga2022_ULD, Erickson2022_ULD, Mattikalli2022_ULD} tackle the problem of bone lesion detection and segmentation. But, a vast majority are for specific lesion types (e.g. lytic/blastic only), while a few of them targeted a multitude of lesion types \cite{Yao2017_mixedSpineMetsDetPETCT,Yao2015_boneMetsCT}. Since different types of bone lesions can appear in various skeletal regions throughout the body, an automated approach for the detection of such lesions is highly desirable; it should account for the intensity variations, and simultaneously be fast and precise.


\begin{figure} [!htb]
\centering
\includegraphics[scale=0.25]{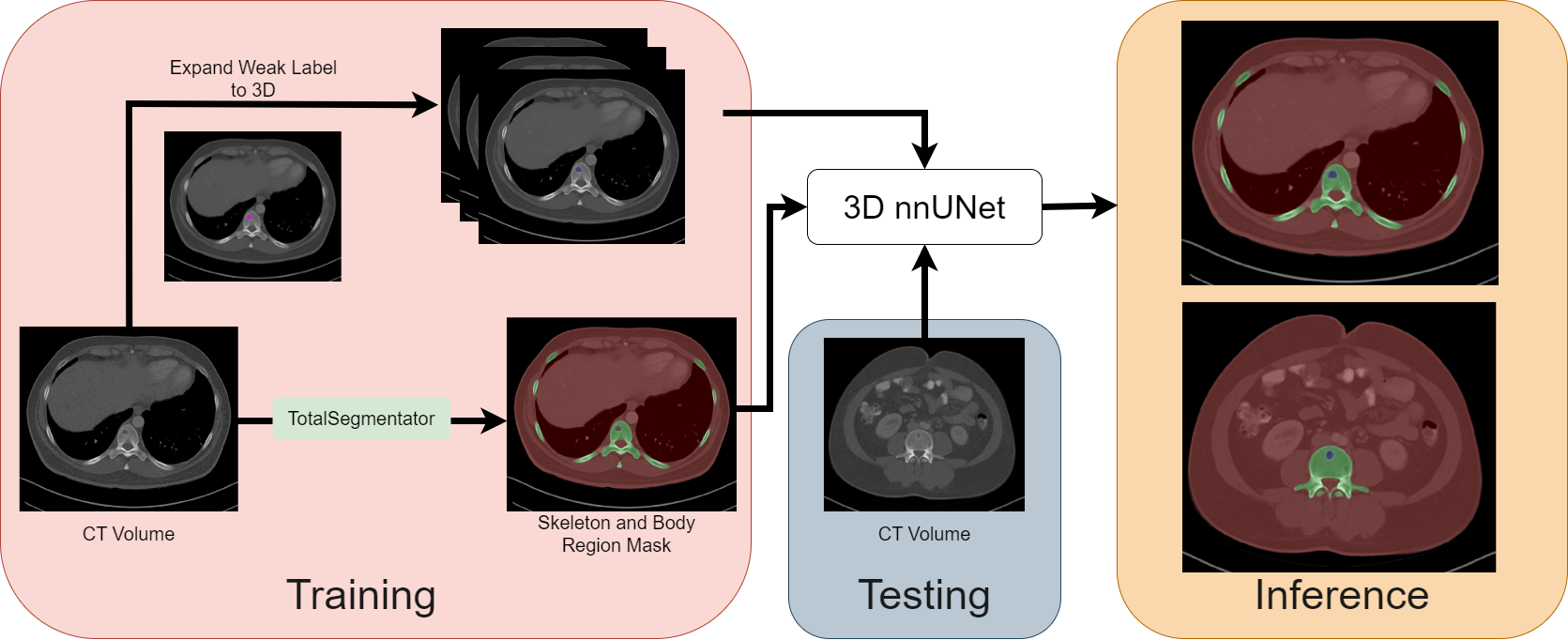}
\medskip
\caption{Framework for the detection of bone lesions via a proxy segmentation task. First, the prospective RECIST measurements (magenta) were used to obtain the lesion mask (blue) via GrabCut. Then, the enclosing bounding box of the lesion mask was extended to create weak 3D masks. Next, the body region (red) and the skeletal region (green) from TotalSegmentator was also merged with the weak 3D masks. Finally, a 3D full-resolution nnUNet model was trained to predict the corresponding regions in the CT volume where the bone lesions were observed. At test time, the model received a 3D CT volume and detected bone lesions in it.}
\label{fig_money}
\end{figure}

As shown in Fig. \ref{fig_money}, we designed an automated pipeline to detect bone lesions (lytic, blastic, and mixed) in CT volumes via a proxy segmentation task. We used the prospectively marked bone lesions in a few 2D slices of CT volumes and converted them into weak 3D segmentation masks. Then, we trained a 3D full-resolution nnUNet model \cite{Isensee_2021} with the 3D masks to detect these lesions by segmenting them. Our model detected bone lesions in CT with high precision (96.7\%), and our results indicated that the partial and incomplete annotations were sufficient to train the nnUNet model for this task. To our knowledge, we are the first to attempt the direct detection of bone lesions (lytic, blastic, and mixed) in CT via a proxy segmentation task.




\section{METHODS}
\label{sec:methods}

\subsection{Dataset}
\label{sec:ds}
The Picture Archiving and Communication System (PACS) at the NIH Clinical Center was queried for patients who underwent CT imaging between 1999 and 2017. Initially, 10,594 studies were collected from 4427 patients \cite{Yan2018_DeepLesion}, and they contained various contrast phases (arterial, portal-venous, delayed, and nephrographic). The scans had varying thicknesses ranging from 1 to 5mm, and the voxel spacings ranged between 0.45 and 0.98 mm and of varying reconstructions (lung, soft tissue, bone). A total of 475 bone lesions were found \cite{Yan2018_DeepLesion} in the 294 CT series from 280 studies of 182 patients. In clinical practice, radiologists prospectively measure lesions with the RECIST measurements; it is made up of two perpendicular lines that denote the extent of a lesion in a CT slice. Certain patients underwent repeat CT studies. The data used in this work was divided on a patient-level into training ($\sim$62\%, 113 patients, 167 studies, 179 series, 279 lesions) and test ($\sim$38\%, 69 patients, 113 studies, 115 series, 196 lesions) splits. The train split had only single patient visits, while patients with $\geq$ 1 visit were filtered into the test split. 

\subsection{Generation of Ground Truth Data}
\label{sec:methods_gt_generation}

\noindent
\textbf{Weak 3D Annotations.} Similar to prior work \cite{Zlocha2019}, we used GrabCut \cite{Rother2004_grabCut} to segment the lesion by exploiting these prospective annotations. First, the CT volume was windowed with a window center and width of [50, 450] HU. Next, a bounding box formed from the measurements defined the maximum search region for the lesion. The interior of this box denoted the probable foreground, while pixels outside it were set to definite background. Then, the endpoints of the measurements on a slice were connected to create a quadrilateral, and its interior was set to definite foreground. Using these definitions, GrabCut was able to segment the lesion, and the segmentations were manually verified (and corrected if necessary). But, the associated RECIST measurements for the adjacent slices were unavailable.

To mitigate this and generate the weak 3D masks, only the bounding box formed from the RECIST measurement was extended to the slices above and below the current slice, and pixels inside this box were set as the segmentation. Our hypothesis behind this strategy was that the model would learn from partial and incomplete annotations since the true lesion extent along the z-axis was unknown. Thus, the weak 3D mask comprised of a 3-slice segmentation mask with the center slice having precise lesion delineations and the adjacent slices having bounding boxes that defined a coarse lesion location. Note that the mask was only made for prospectively marked lesions in the training split and not all lesions. The test split was fully reviewed by a medical school graduate applying for diagnostic radiology residency (A.S.) and confounding cases were reviewed by a senior board-certified radiologist with 30+ years of experience (R.M.S.).


\noindent
\textbf{Skeletal and Body Region Segmentation.} Since bone lesions can be small or large, a nnUNet model trained directly with the weak 3D masks will have degraded performance due to a class imbalance between the foreground and background (all other regions) class. To combat this issue, we also utilized the segmentations of the skeletal and body regions as shown in Fig. \ref{fig_money}. These segmentations were generated by TotalSegmentator \cite{Wasserthal2023}, which is a tool that produced segmentation masks for various organs and structures in CT. The skeletal region segmentation comprised of the spine, left and right ribs, scapulas, clavicles, femurs, hip bones, sacrum, and sternum. The weak 3D annotations were merged with the body and skeletal region segmentations for training.

\subsection{Detection of Bone Lesions using nnUNet} 

The nnUNet \cite{Isensee_2021} is a self-configuring segmentation framework that can be adapted to different datasets and modalities, such as CT. It automatically determined the optimal hyper-parameters for training a segmentation model and learned to segment target structures of interest. We trained a 3D full-resolution nnUNet for detecting bone lesions via a proxy segmentation task. During training, our pipeline took a CT volume and the ground-truth masks from Sec. \ref{sec:methods_gt_generation} as input. nnUNet learned to generate a segmentation for the CT volume and iteratively refined it via a loss function that measured the segmentation error through the overlap between the prediction and ground-truth. At inference time, nnUNet predicted the segmentation mask for an input CT volume. The mask included the bone lesions present in the volume along with the skeletal and body region masks. 

\begin{figure}[!tb]
    \centering
    
    \begin{subfigure}[t]{0.245\textwidth}
        \includegraphics[width=\textwidth]{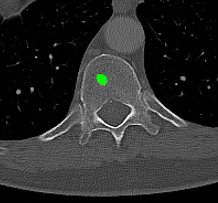}
    \end{subfigure}
    \hfill
    \begin{subfigure}[t]{0.245\textwidth}
        \includegraphics[width=\textwidth]{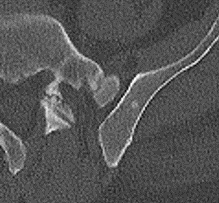}
    \end{subfigure}
    \hfill
    \begin{subfigure}[t]{0.245\textwidth}
        \includegraphics[width=\textwidth]{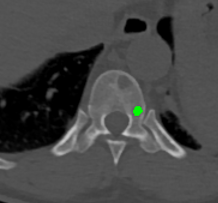}
    \end{subfigure}
    \hfill
    \begin{subfigure}[t]{0.245\textwidth}
        \includegraphics[width=\textwidth]{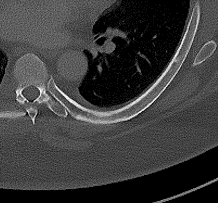}
    \end{subfigure}

    \smallskip

    \begin{subfigure}[t]{0.245\textwidth}
        \includegraphics[width=\textwidth]{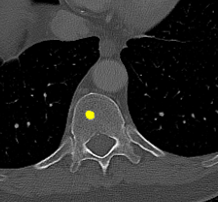}
        \caption{}
    \end{subfigure}
    \hfill
    \begin{subfigure}[t]{0.245\textwidth}
        \includegraphics[width=\textwidth]{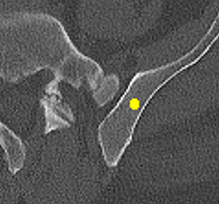}
        \caption{}
    \end{subfigure}
    \hfill
    \begin{subfigure}[t]{0.245\textwidth}
        \includegraphics[width=\textwidth]{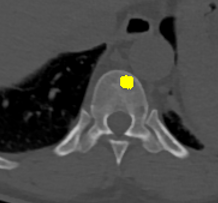}
        \caption{}
    \end{subfigure}
    \hfill
    \begin{subfigure}[t]{0.245\textwidth}
        \includegraphics[width=\textwidth]{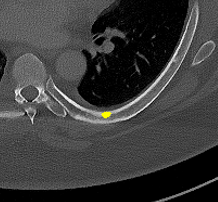}
        \caption{}
    \end{subfigure}

    \medskip

    \caption{Qualitative results of bone lesion detection. Each column shows a cropped slice with the top row displaying ground-truth (green) and the bottom row showing nnUNet predictions (yellow). (a) shows a TP osteolytic lesion; (b) shows the prediction of a real osteoblastic lesion that was not prospectively marked as ground-truth (before review), and after GT review, it was correctly considered as ground-truth; (c) shows an osteolytic lesion not annotated in ground-truth before review and correctly marked afterwards. A FN adjacent to it is also shown. (d) shows a rare FP on the cortex of a rib.}

    \label{fig_results}
\end{figure}

\section{Experiments and Results}
\label{sec:exp_results}

\textbf{Implementation.} Our 3D nnUNet model was trained using 5-fold cross-validation with different initializations of trainable parameters for a total of 1000 epochs. The loss function used by the model was an equally weighted combination of robust cross-entropy and memory efficient soft Dice losses. It was optimized using the Stochastic Gradient Descent (SGD) optimizer with an initial learning rate of 0.01 and a batch size of 2. Each CT volume in the test split was passed to the model from each fold, and predictions from five folds were ensembled together by nnUNet. Each fold was trained on the NIH HPC Biowulf Cluster with an NVIDIA V100 GPU with 32GB RAM.

\noindent
\textbf{Metrics.} We used precision and recall as the evaluation metrics to quantify the performance of our model. True positives (TPs) were those ground truth lesions that overlapped with predicted lesions. False positives (FPs) were those predictions that did not overlap with any ground truth. Finally, false negatives (FNs) were ground truth lesions that were missed by our model.


\noindent
\textbf{Results.} Qualitative and quantitative results of our model are presented in Fig. \ref{fig_results} and Table \ref{table_results} respectively. Our model achieved a precision of 96.7\% and a recall of 47.3\% for detection of bone lesions (lytic, blastic, and mixed). We observed that the model was very confident in its predictions as demonstrated by the vast number of actual lesion predictions (TPs) compared against the FPs. Despite the high precision, the model was not very sensitive to the detection of bone lesions due to the large number of FNs.

\begin{table*}[!htb]
\centering\fontsize{10}{12}\selectfont 
\setlength{\tabcolsep}{7pt} 
\setcellgapes{3pt}\makegapedcells 
\caption{Results of bone lesion detection with 3D full-resolution nnUNet.} 
\begin{adjustbox}{max width=\textwidth}
\begin{tabular}{l|c|c|c|c|c|c|c}
\hline
 & \# Scans & \# GT & \# TP & \# FP & \# FN & Precision & Recall \\
\hline 
After GT review & 115 & 793 & 375 & 13 & 418 & 96.7 & 47.3 \\
Before GT review & 115 & 196 & 99 & 289 & 98 & 25.6 & 50.2 \\
\hline
\end{tabular}
\end{adjustbox}
\label{table_results}
\end{table*}



\section{DISCUSSION AND CONCLUSION}
\label{sec:discussion}

In this pilot work, we proposed to automatically detect bone lesions in CT volumes via a proxy segmentation task using the 3D full-resolution nnUNet model. Despite the partial and incomplete 3D annotations used to train the model, it was very precise in its predictions (96.7\%). Due to the nature of the weak 3D annotations, the middle slices of the predictions provided better delineations of lesions in contrast to the adjacent slices (which were only coarse bounding boxes). These results were obtained after a thorough qualitative review of the model's predictions. Initially, the test split also comprised of only prospectively marked lesions with other potential lesions being left unmarked. We noticed that the model captured lesions that were not originally annotated by the reading radiologist. Despite the vast majority of the predictions that were considered as FPs were, in fact, real lesions (TPs), the precision and recall languished at 25.6\% and 50.2\% respectively as shown in Table \ref{table_results} (Before GT review). The discordance between the initial quantitative and qualitative results spurred us towards a full review of our ground-truth annotations in the test split. Post completion of the review, the recomputed metrics resulted in a significant increase in the precision of our model, which was in line with the observed qualitative results.

There are several limitations to this work. First, the model was not trained on fully annotated training data, and this affected the detection performance of large lesions since there were few examples of large lesions ($\geq$ 3cm) during training. To illustrate this problem of incomplete annotations, there were 196 original prospective RECIST measurements in the test split, but we found 793 confirmed lesions after GT review (4$\times$ fold increase). Next, the RECIST measurements were only made on one slice and unavailable for the adjacent slices, thus the 3D annotations were incomplete and partial. Additionally, we do not report Dice scores in this work because the training data comprised of weak annotations and it was not possible to compute accurate Dice scores. Prior works obtained higher sensitivities $\geq$ 75\% for lesion detection, but this came at the expense of a large number of FPs. Such computer-aided detection systems would overwhelm the radiologist with FPs when used as second reader. Thus, we argue that a model should be both precise in detecting existing lesions and sensitive enough to capture new lesions. However, we hypothesize that the number of FNs come from incomplete training annotations. During training, the model receives full CT volumes, all of which have numerous unannotated lesions. This undoubtedly impacted it at inference time as the model will have learned many real lesions are background. Addressing the missing annotations issue will be imperative for precise and sensitive bone lesion detection. However, it would require substantial efforts to create full 3D voxel-level annotations of all bone lesions in all CT volumes. Our pilot results only represent the first step in detecting bone lesions with high precision using partial and incomplete 3D annotations.

\acknowledgments 
This work was supported by the Intramural Research Programs of the NIH Clinical Center. We utilized the computational resources of the \href{http://hpc.nih.gov}{NIH HPC Biowulf cluster}.

\clearpage
\bibliography{report} 
\bibliographystyle{spiebib} 




\end{document}